# A Novel Hybrid Algorithm for Permutation Flow Shop Scheduling

Sandeep Kumar, Pooja Jadon

*Computer Science Department, Jagannath University*
*Chaksu Campus, Jaipur*

*Computer Science Department, Jagannath University*

**Abstract**— In the present scenario the recent engineering and industrial built-up units are facing hodgepodge of problems in a lot of aspects such as machining time, electricity, man power, raw material and customer's constraints. The job-shop scheduling is one of the most significant industrial behaviours, particularly in manufacturing planning. This paper proposes the permutation flow shop sequencing problem with the objective of makespan minimization using the new modified proposed method of johnson's algorithm as well as the gupta's heuristic algorithm. This paper involves the determination of the order of processing of n jobs in m machines. Although since the problem is known to be np-hard for three or more machines, that produces near optimal solution of the given problem. The proposed method is very simple and easy to understand followed by a numerical illustration is given.

## I. INTRODUCTION

Scheduling is the allocation of resources (e.g. machines) to tasks (e.g. jobs) in order to ensure the completion of these tasks in a reasonable amount of time. The selection of appropriate job is selected using some methods. There are limited jobs can be executed over the processor on the behalf of availability of resources. The Uni-processor or multiprocessor scheduling is to be done according to used processor.

The major task is to find out the optimal solution for the execution of jobs on different machines or processors. The number of processor or jobs may be different from each other. So it is a critical task to do.

**Job shop** is a work location where a number of general purpose work stations or machines exist and are used to perform a variety of jobs.

**Job shop scheduling** is also known as job shop problem. It is an optimization predicament in which resources are
 allocated to ideal jobs at particular times. State n jobs $J_1$, $J_2$,…. $J_n$ and each job consists a chain of operations with different sizes, which scheduled on m indistinguishable machines (m > 2), while demanding to minimize the makespan. Each machine can process at most one operation at a time. In any process the preemption of an operation is not acceptable.

**Flow shop scheduling** is a special case of job shop scheduling in which all operations are having strict order to be performed all jobs. Flow shop scheduling problems are scheduling problems in which the flow control shall enable an appropriate sequencing for each job and for processing on a set of machines or with other resources 1,2,...,m in compliance with given processing orders. It maintains the constant flow of processing jobs is preferred with a minimum of idle time and a minimum of waiting time.

**Flowshop Scheduling:** Flowshop Scheduling determines an optimum sequence of n jobs to be processed on m machines in the same order i.e. every job must be processed on machines 1,2,...,m in this same order.

The flowshop scheduling problem is a production problem where a set of *n* jobs have to be processed with identical flow patterns on *m* machines. When the sequence of jobs processing on all machines is the same we have the permutation flowshop sequencing production environment. We study the flow-shop problems considering the following assumptions:

The operation processing times on the machines are known, fixed and some of them may be zero if some job is not pro-cessed on a machine.

Set-up times are included in the processing times and they are independent of the job position in the sequence of jobs.

At a time, every job is processed on only one machine, and every machine processes only one job.

The job operations on the machines may not be pre-empted [4].

A flowshop consists of *M* machines in series and *N* different jobs available for processing at time zero. Each machine can handle only one job at a time. Each job is continuously processed on *M* available machines in the same technological order. This makes it different from a typical jobshop problem. The processing time including a setup time of job performed on the machine depends on the amount of constrained resources allocated to these jobs. The processing time of job *i* on machine *j* is denoted by $t_{ij}$ for $i = 1, 2, …, N$ and $j = 1, 2, …, M$ [12].

The heuristic algorithms are more efficient and economical of getting a practical solution, though it sometimes cannot reach the optimum [12].

In the flow shop scheduling problem, n jobs are to be processed on m machines. The order of the machines is fixed. We assume that a machine processes one job at a time and a job is processed on one machine at a time without pre-emption. The problem of flow shop scheduling is NP-complete. In general, there are (n!) m different schedules of jobs [9].

The time required to complete the process of all the operation is called as Total processing time or makespan. There are some assumptions considering the JSP problem.

Each job consists of a finite number of operations.

The processing time for each operation has been determined.

Machine Sequence is defined for each job.

There is a pre-defined sequence of operations that has to be maintained to complete each job.

Job does not visit the same machine twice.

Set up time included in the Processing time.

A machine can process only one job at a time.

No machine can deal with more than one type of task.

Operations cannot be interrupted.

Neither release time nor due dates are specified.

Each job should be processed through the machines in a particular order or also known as technological Constraints.[10]

The algorithm of Johnson is a classic method which solves to optimum the problem of ordering n jobs on two machines, in a polynomial time. If there are n jobs on three machines, then the problems become NP-complete (i.e. cannot be solved optimally in polynomial time) and the Johnson's algorithm can be applied only for some few particular cases that obey some primary conditions [11]. Recently J.F. Gonçalves et al. [15] wished-for a local search method for the job-shop scheduling problem based on a new neighborhood, and it generate schedule decoding the chromosome supplied by the genetic algorithm. A Scatter Search (SS) based algorithm presented by [16] to solve job shop scheduling problems. It considers the availability constraints in a fuzzy job shop scheduling problem. It considers that a machine can be unavailable for some reason like maintenance, repairing or due to sudden breakdown. It minimizes the tardiness and earliness in fuzzy job shop scheduling problem. S. Sundar et al. [17] wished-for a flexible job-shop scheduling problem with no-wait constraint (FJSPNW). It combines characteristics of two different job shop scheduling problems, flexible and no-wait job-shop scheduling problem. R. Zhang et al. [18] wished-for a hybrid differential evolution (DE) method to solve the job shop scheduling problem with random processing times with the target of minimizing the expected total tardiness. X Zhang et al. [19] addressed an intuitionist fuzzy set Job Shop Scheduling Problem. R Kumari et al. proposed a fuzzified job shop scheduling algorithm based on gupta's heuristics.

Rest of the paper is organized as follows: Permutation Flowshop Scheduling in section 2. In section 3, job shop scheduling problem is explained. Section 4 introduces gupta's heuristics and next section discusses proposed hybrid job shop scheduling algorithm. In section 6, performance of the proposed strategy is analyzed. Finally, in section 7, paper is concluded.

## II. PERMUTATION FLOWSHOP SCHEDULING

Permutation Flowshop Scheduling is a special case of FSPs where same job sequence is followed in all machines i.e. processing order of the jobs on the machines is the same for every machine [1].

The permutation flow shop scheduling problem (PFSP) is a production problem for finding the best sequence of jobs that to be processed by machines in order to minimize the given objective function. This case can be found in manufacturing facilities where the jobs (parts) are moved from machine to machine by material handling devices with no passing is allowed. The problem has been proved to be strongly NP-complete and its total number of possible schedules (sequence) is for jobs. The total flow time and the makespan are of important performance measures, which lead to rapid turn-around of jobs and minimization of in-process inventory [3].

The permutation flow shop scheduling problem is considered as NP-Hard problem. A scheduling problem is NP-hard in the ordinary sense if

Partition (or a similar problem) can be reduced to this problem with a polynomial time algorithm and

There is an algorithm with pseudo polynomial time complexity that solves the scheduling problem [6].

It is in the set of NP (nondeterministic polynomial time) problems: Any given solution to L can be verified quickly (in polynomial time).

It is also in the set of NP-hard problems: Any NP problem can be converted into L by a transformation of the inputs in polynomial time. Although any given solution to such a problem can be verified quickly, there is no known efficient way to locate a solution in the first place; indeed, the most notable characteristic of NP-complete problems is that no fast solution to them is known. That is, the time required to solve the problem using any currently known algorithm increases very quickly as the size of the problem grows.

Scheduling is the method by which threads, processes or data flows are given access to system resources (e.g. processor time, communications bandwidth).The need for a scheduling algorithm arises from the requirement for most modern systems to perform multitasking (execute more than one process at a time) and multiplexing (transmit multiple flows simultaneously).

The scheduler is concerned mainly with:

Throughput - The total number of processes that complete their execution per time unit.

CPU or Processor Utilization: We want to keep the CPU as busy as possible [2].

Latency, specifically:

Turnaround time - total time between submission of a process and its completion. Or in other words turnaround time is the sum of the periods spent waiting to get into memory [2].

Response time - amount of time it takes from when a request was submitted until the first response is produced.

Fairness / Waiting Time - Equal CPU time to each process (or more generally appropriate times according to each process' priority). It is the time for which the process remains in the ready queue [2].

## III. JOB SHOP SCHEDULING PROBLEM

The flow shop problem can be formulated as follows. Each of n jobs from the job set $i = \{1, 2,...., n\}$, for $n > 1$, has to be processed on m machines $1, 2, ...., m$ in the order given by the indexing of the machines. Thus job $j, j \in J$, consists of sequence of m operations; each of them corresponding to the processing of job j on machine i during an uninterrupted processing time $p_{ij} \geq 0$. It is assumed that a zero processing time on a machine corresponds to a job performed by that machine in an infinitesimal time. Machine $j$, $j = 1, 2, ....,m$, can execute at most one job at a time, and it is assumed that each machine processes the job in the same order . The objective is to find a sequence for the processing of the jobs on the machines so that the total completion time or makespan of the schedule ($C_{max}$) is minimized. The processing times needed for the jobs on the machines are denoted as $p_{ij}$, where $i = 1,…,n$ and $j = 1, . . . ,m$; these times are fixed, known in advance and non-negative.

There are several assumptions that are made regarding this problem:

Each job $i$ can be processed at most on one machine $j$ at the same time.

Each machine $m$ can process only one job $i$ at a time.

No preemption is allowed, i.e. the processing of a job $I$ on a machine $j$ cannot be interrupted.

The set-up times of the jobs on machines are included in the processing times.

The machines are continuously available.

In-process inventory is allowed. If the next machine on the sequence needed by a job is not available, the job can wait and join the queue at that machine [5].

## IV. THE GUPTA'S HEURISTIC ALGORITHM

Gupta anticipated a heuristic technique [13] to achieve an almost minimum makespan. In the Gupta heuristic algorithm all the jobs are divided into two groups by comparing the dispensation times of the first machine and the last machine in each job. For every group, calculate the sum of processing times of any two adjacent tasks in a job and find the minimum processing time, and then schedules the jobs in sorting order according to their minimum summed processing times.

Johnson's algorithm is basically used for only two machines [14], but the idea of Gupta algorithm is applicable for more than two machines [13]. This algorithm state an m machines, a set of n independent jobs with a chain of operations that must be executed in the same sequence on each machine. Gupta proposed the following heuristic algorithm to solve it in polynomial time [13]. The below mentioned Algorithm outline Gupta's heuristic.

Input: A set of $n$ jobs, each having $m$ ($m > 2$) tasks executed respectively on each of $m$ machines.

Output: A schedule with a nearly minimum completion time of the last job.

*Step 1:* Form the group of jobs $U$ that take less time on the first machine than on the last
such that $U = \{ i \mid t_{1i} < t_{mi}\}$.

*Step 2:* Form the group of jobs $V$ that take less time on the last machine than on the
first such that $V = \{ j \mid t_{mj} \leqq t_{1j}\}$.

*Step 3:* For each job $J_i$ in $U$, find the minimum of ($t_{kj} + t_{(k+1)j}$) for $k = 1$ to $m$-1;
  restated set:
  $\pi_i = \min(t_{ki} + t_{(k+1)i})$ for k=1 to m-1

*Step 4:* For each job $Jj$ in $V$, find the minimum of ($t_{kj} + t_{(k+1)j}$) for $k = 1$ to $m$-1;
  Restated set:
  $\pi_j = \min(t_{kj} + t_{(k+1)j})$ for k=1 to m-1

*Step 5:* Sort the jobs in $U$ in ascending order of $\pi_i$'s; if two or more jobs have the same value of $\pi_i$, sort them in an arbitrary order.

*Step 6:* Sort the jobs in $V$ in descending order of $\pi_j$'s; if two or more jobs have the same value of $\pi_j$, sort them in an arbitrary order.

*Step 7:* Schedule the jobs on the machines in the sorted order of $U$, then in the sorted order of $V$.[8]

## V. PROPOSED NEW MODIFIED ALGORITHM FOR PERMUTATION FLOW SHOP JOB SCHEDULING

The anticipated job shop scheduling algorithm is a modification in gupta's heuristic. This algorithm is amalgamation of Gupta's heuristic and Johnson's algorithm. It deals with some fuzzy logic rules and these rules are based on each operation time of a job. It calculates a new value of execution with the help of operation time of each job for combination of two machines (like, machine 1 and machine 2, machine 2 and machine 3 and so on). This paper

Input: A set of $n$ jobs, each having $m$ ($m > 2$) tasks executed respectively on each of m *machines*.

Output: A schedule with a nearly minimum completion time of the last job.

*Step 1:* Maximum processing time on machine1 should be greater than or equal to mimimum processing time on machine m2 m3 m-1.

*Step 2:* Minimum processing time on machine m should be greater than or equal to maximum processing time on machine m2 m3 m-1.

*Step 3:* If one of these above condition is met then find out the shortest processing time.

*Step 4:* Form two hypothetical machines on which we perform the summed up of job's processing time.

*Step 5:* Form the group of jobs $U$ that take less time on the first machine than on the last
such that $U = \{ i \mid t_{1i} < t_{mi}\}$.

*Step 6:* Form the group of jobs $V$ that take less time on the last machine than on the
first such that $V = \{ j \mid t_{mj} \leqq t_{1j}\}$.

*Step 7*: Merge these two group by taking the first two jobs and schedule them in order to minimize the partial makespan as if there were only these two jobs

*Step 8*: Calculate the make-span time for the sequence obtained in step 7.

*Step 9*: Find the ordered pair of jobs from *'job list'*, which corresponds at a minimum makespan.

*Step 10:* Set k = 3

Select the $k^{th}$ job from the sorted list and insert it into $k$ possible positions of the best partial sequence. Among the $k$ partial sequences, select the best one with the least makespan.

Set k = k + 1

Example: Consider 7 jobs, 4 machine flow shop problem with processing time as shown in table 1.

TABLE I: PROBLEM

| JOBS | $M_1$ | $M_2$ | $M_3$ | $M_4$ |
|---|---|---|---|---|
| $J_1$ | 3 | 1 | 4 | 12 |
| $J_2$ | 8 | 0 | 5 | 15 |
| $J_3$ | 11 | 3 | 8 | 10 |
| $J_4$ | 4 | 7 | 3 | 8 |
| $J_5$ | 5 | 5 | 1 | 10 |
| $J_6$ | 10 | 2 | 0 | 13 |
| $J_7$ | 2 | 5 | 6 | 9 |

Makespan calculation using GUPTA's Algorithm:
Group X= {$J_1, J_2, J_4, J_5, J_7$}
Group Y= {$J_3, J_6$}
$J_1$= Min {(3+1, 1+4, 4+12)} = Min (4, 5, 16) = 4
$J_2$= Min {(8+0, 0+5, 5+15)} = Min (8, 5, 20) = 5
$J_3$= Min {(11+3, 3+8, 8+10)} = Min(14,11,18) = 11
$J_4$= Min {(4+7, 7+3, 3+8)} = Min(11, 10, 11) = 10
$J_5$= Min {(5+5, 5+1, 1+10)} = Min(10, 6, 11)= 6
$J_6$= Min {(10+2, 2+0, 0+13)} = Min(12, 2, 13)= 2
$J_7$= Min {(2+5, 5+6, 6+9)} = Min(7, 11, 15)= 7
Ordered Group in Ascending X= {$J_1, J_2, J_5, J_7, J_4$}
Ordered Group in Descending Y= {$J_3, J_6$}
The Total Sequence is = {$J_1, J_2, J_5, J_7, J_4, J_3, J_6$}
Makespan for that Sequence is = 85

Table II: Makespan calculation using GUPTA's Algorithm

| JOBS | M1 | M2 | M3 | M4 |
|---|---|---|---|---|
| J1 | 0-3 | 3-4 | 4-8 | 8-20 |
| J2 | 3-11 | 11-11 | 11-16 | 20-35 |
| J5 | 11-16 | 16-21 | 21-22 | 35-45 |
| J7 | 16-18 | 21-26 | 26-32 | 45-54 |
| J4 | 18-22 | 26-33 | 33-36 | 54-62 |
| J3 | 22-33 | 33-36 | 36-44 | 62-72 |
| J6 | 33-44 | 44-46 | 46-46 | 72-85 |

Makespan calculation using a novel hybrid permutation flow shop scheduling
Maximum Processing time on machine1 should be greater than or equal to maximum processing time on machine $m_2, m_3,…m-1$ → 11 >= 0

Minimum Processing time on machine m should be greater than aor equal to the maximum processing time on machine $m_2, m_3…m-1$ → 8>=8
Condition satisfies then we introduce two hypothetical machines X and Y respectively.

| JOBS | X | Y |
|---|---|---|
| $J_1$ | 8 | 17 |
| $J_2$ | 13 | 20 |
| $J_3$ | 22 | 21 |
| $J_4$ | 14 | 18 |
| $J_5$ | 11 | 16 |
| $J_6$ | 12 | 15 |
| $J_7$ | 13 | 20 |

Summed up the processing time first three machines namely $M_1+M_2+M_3$= X
Summed up the processing time last three machines namely $M_2+M_3+M_4$= Y
Ordered Group in Ascending X= {J1, J5, J6, J7, J2, J4}
Ordered Group in Descending Y= {J3}
Now, therefore we have to find out the ordered sequence of the unscheduled jobs according their processing time such that
{$J_1, J_5, J_6, J_7, J_2, J_4, J_3$}

Table III: Makespan calculation using a novel hybrid permutation flow shop scheduling

| JOBS | $M_1$ | $M_2$ | $M_3$ | $M_4$ |
|---|---|---|---|---|
| $J_1$ | 0-3 | 3-4 | 4-8 | 8-20 |
| $J_5$ | 3-7 | 7-14 | 14-17 | 20-28 |
| $J_6$ | 7-17 | 17-19 | 19-19 | 28-41 |
| $J_7$ | 17-19 | 19-24 | 24-30 | 41-50 |
| $J_2$ | 19-27 | 27-27 | 30-35 | 50-65 |
| $J_4$ | 27-31 | 31-38 | 38-41 | 65-73 |
| $J_3$ | 31-42 | 42-45 | 45-53 | 73-83 |

Makespan for this generated sequence is = 83
Thus it can be observed from solutions that proposed novel hybrid permutation job shop scheduling algorithm is better than existing algorithms.

## VI. CONCLUSION

This paper proposed a new hybrid job shop scheduling algorithm for more than two machines. This algorithm provides far better result than Gupta's heuristic. It provides minimum makespan as well as better partial makespan. Proposed algorithm also provides more than one choice for optimum result. We get a sequence with makespan less than Gupta's heuristic. So, the proposed new modified job shop scheduling algorithm able to reduce the makespan.